\documentclass[11pt]{article}
\usepackage{latexsym,graphicx,a4}
\usepackage{amsmath,amssymb}
\usepackage{amsfonts}
\usepackage{graphics,epsfig}
\begin{document}
\begin{center}
{\Large\bf The BSW Impact Picture 30 Years After}\footnote{Contribution
to Diffraction 2008, International Workshop on Diffraction
in High Energy Physics, La Londe-les-Maures, September 9-14 2008.}
\vskip1.4cm

{\bf Claude BOURRELY}
\vskip 0.4cm
Universit\'e de la M\'editerran\'ee, Aix-Marseille II,
\\Facult\'e des Sciences - Luminy, 163,  Avenue de Luminy,
\\ 13288 Marseille cedex 9, France
\vskip 0.4cm

\bf{Abstract}
\end{center}
Thirty years ago Bourrely-Soffer-Wu (BSW) investigated an impact picture
description of $p-p$ and $\bar p -p$ elastic scattering with success.
For this anniversary, a short review of the main features of this picture
will be presented with its evolution over the years. The future
experiments at the LHC collider will provide decisive tests for the
impact picture at high energy. 
\vskip 1.0cm

In the year 1978, BSW proposed an impact parameter description of the
high energy behavior of elastic ${p-p}$ and $\bar p -p$ scattering \cite{r8}. 
A main feature
was the peculiar form of the energy dependence of the pomeron,
based on the results of Cheng and Wu \cite{r1,r2}
derived from the high-energy behavior of
quantum field theory. Later, in 1984, with the
advent of ${\bar p - p}$ experiments, a more complete analysis was 
performed \cite{r9,r10}.
I will present a recent update of the results which are focused on the
LHC energy domain.
An extension to the elastic processes ${\pi - p}$ and 
${K - p}$ can be found in \cite{r16}.
Moreover, we have also shown that under certain assumptions 
${\gamma -p}$ and ${\gamma - \gamma}$
total cross sections can be predicted \cite{r11}.

In the impact-picture representation, the spin-independent scattering 
amplitude\footnote{ Here we neglect the spin-dependent
amplitude which was considered in Refs.~\cite{r8,r13} 
for the description of polarizations and spin correlation parameters.}, 
for $p p$ and $\bar p p$ elastic scattering, reads as
\begin{equation}\label{ampli}
a(s,t) = \frac{is}{2\pi}\int e^{-i\mathbf{q}\cdot\mathbf{b}} (1 - 
e^{-\Omega_0(s,\mathbf{b})})  d\mathbf{b} \ ,
\end{equation}
where $\bf q$ is the momentum transfer ($t={-\bf q}^2$) and 
$\Omega_0(s,\mathbf{b})$ is defined to be the opaqueness at impact parameter 
$\bf b$ and at a given energy $s$. We take 
\begin{equation}\label{opac}
\Omega_0(s,\mathbf{b}) = S_0(s)F(\mathbf{b}^2)+ R_0(s,\mathbf{b}) \ ,
\end{equation}
the first term is associated with the pomeron exchange, which generates 
the diffractive component of the scattering and the second term is 
the Regge background.
The Pomeron energy dependence is given by the crossing symmetric expression 
\cite{r1,r2} 
\begin{equation}\label{energ}
S_0(s) = \frac{s^c}{(\ln s)^{c'}} + \frac{u^c}{(\ln u)^{c'}} \ ,
\end{equation}
where $u$ is the third Mandelstam variable.
The choice one makes for $F(\mathbf{b}^2)$ is crucial and we take the Bessel 
transform of
\begin{equation}\label{formf}
\tilde F(t) = f[G(t)]^2 \frac{a^2 + t}{a^2 -t} \ ,
\end{equation}
where $G(t)$ stands for the proton "nuclear form factor", parametrized like 
the electromagnetic form factor, as a two poles, 
\begin{equation}\label{fgt}
G(t) = \frac{1 }{(1 - t/m_1^2)(1 - t/m_2^2)} \ .
\end{equation}
The slowly varying function occuring in Eq.(\ref{formf}), reflects the 
approximate proportionality between the charge density and the hadronic matter
distribution inside a proton.
So the pomeron part of the amplitude depends on only {\it six} parameters
$c, c', m_1, m_2, f,$ and $a$. 
The asymptotic energy regime of hadronic interactions are controlled by 
$c$ and $c'$, which will be kept, for all elastic reactions, at
the values obtained in 1984 \cite{r9}, namely
\begin{eqnarray}\label{cc'}
& c=0.1675 \quad c'=0.7479 \quad m_1 = 0.5763\mbox{GeV}^2 \quad
m_2 = 0.5763\mbox{GeV}^2 \nonumber \\
& f = 6.9968 \quad a = 1.9354\mbox{GeV}^2.
\end{eqnarray}
The remaing four parameters are related, more specifically to the reaction 
$pp$ ($\bar p p$) and they will be slightly re-adjusted from 
the use of a new set of data.

We now turn to the Regge background. A generic Regge exchange amplitude 
has an expression of the form
\begin{equation}\label{ampreg}
\tilde R_i(s,t)=C_ie^{b_it} \left[ 1 \pm e^{-i\pi\alpha_i(t)}\right]
(\frac{s}{s_0})^{\alpha_i(t)} \ ,
\end{equation}
where $C_ie^{b_it}$ is the Regge residue, $\pm$ is the signature factor, 
$\alpha_i(t) = \alpha_{0i} + \alpha_i^{'} t $ is 
an effective linear Regge trajectory and $s_0 =1$GeV$^2$.
If we consider the sum over all the allowed Regge trajectories 
$\tilde R_0(s,t)= \sum_i \tilde R_i(s,t)$, the Regge background 
$R_0(s,\mathbf{b})$ in Eq. (\ref{opac}) is the Bessel transform of
$\tilde R_0(s,t)$. In $pp$ ($\bar p p$) elastic scattering,
the allowed Regge exchanges are $A_2$, $\rho$, $\omega$, so the Regge 
background involves several additional parameters which are described
in \cite{r16}.

\begin{figure}[ht]
\vspace*{-2.5ex}
  \begin{minipage}{7.0cm}
  \epsfig{figure=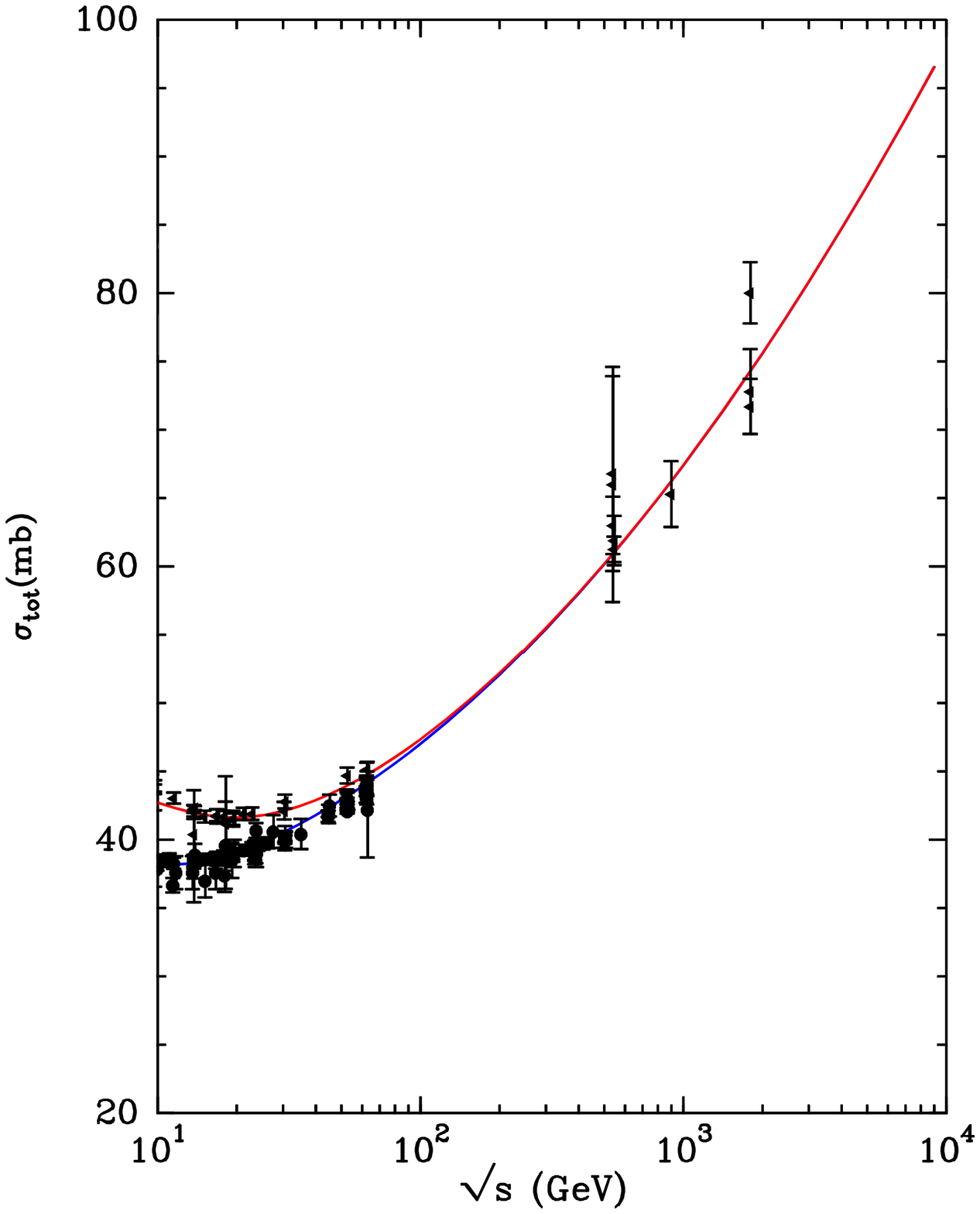,height=8.7cm,width=6.5cm}
  \end{minipage}
    \begin{minipage}{7.0cm}
  \epsfig{figure=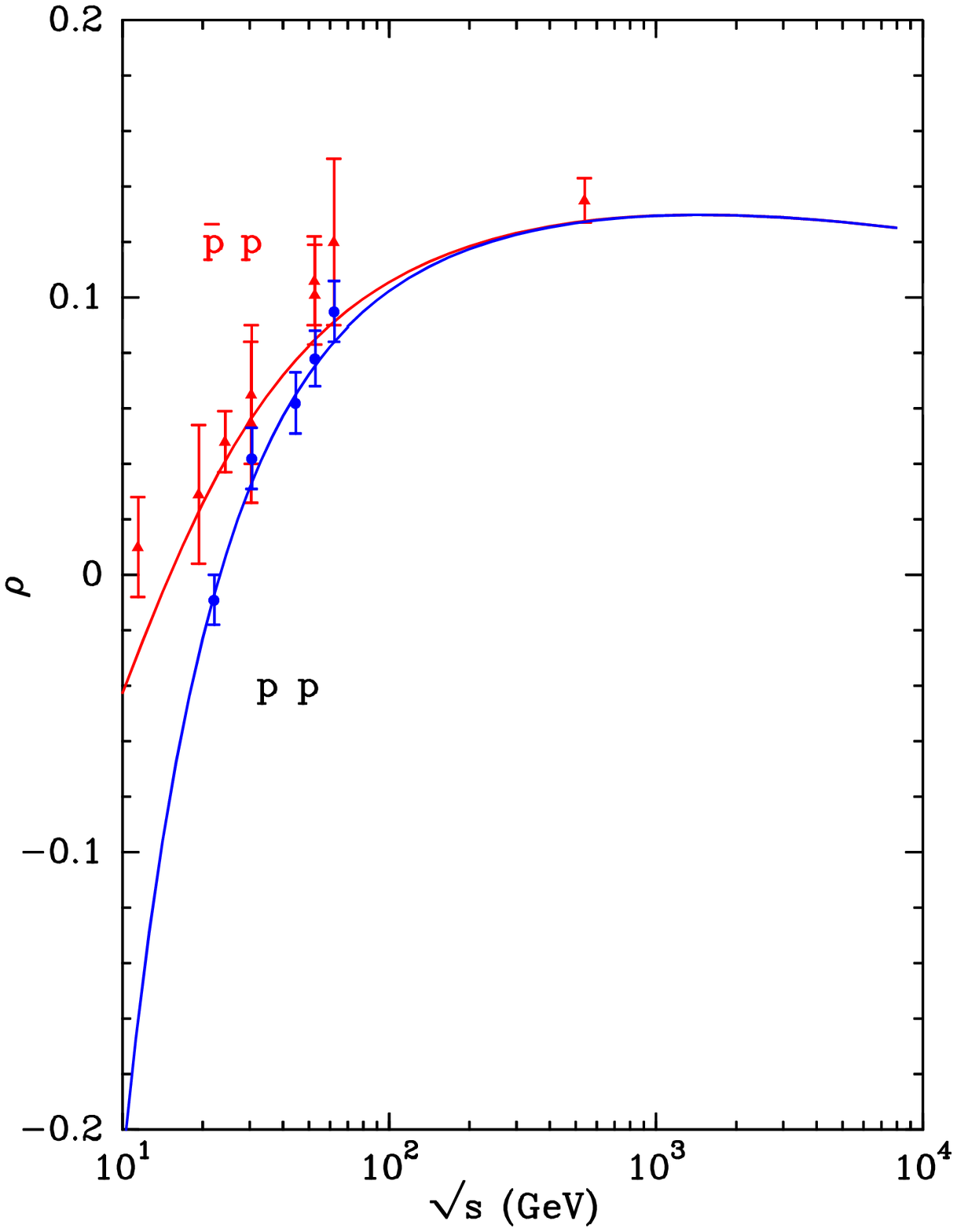,height=8.7cm,width=6.5cm}
    \end{minipage}
\vspace*{-2.0ex}
\caption{
$p p$ and $\bar p p$ elastic scattering,  $\sigma_{tot}$ (left),
 $\rho$ (right) as a function of the energy.}
\label{fi:fig1}
  \begin{minipage}{7.0cm}
  \epsfig{figure=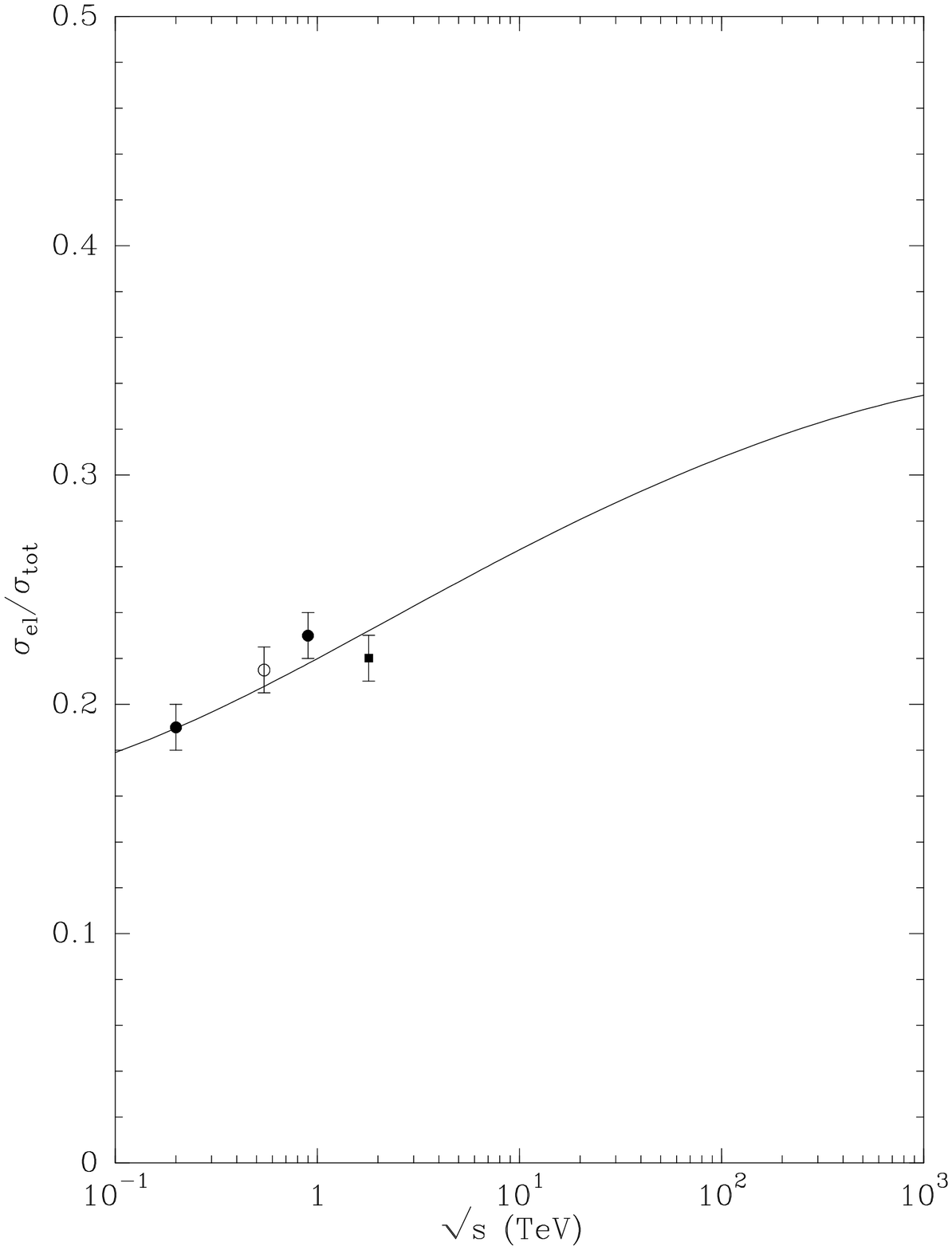,height=8.5cm,width=6.3cm}
  \end{minipage}
    \begin{minipage}{7.0cm}
  \epsfig{figure=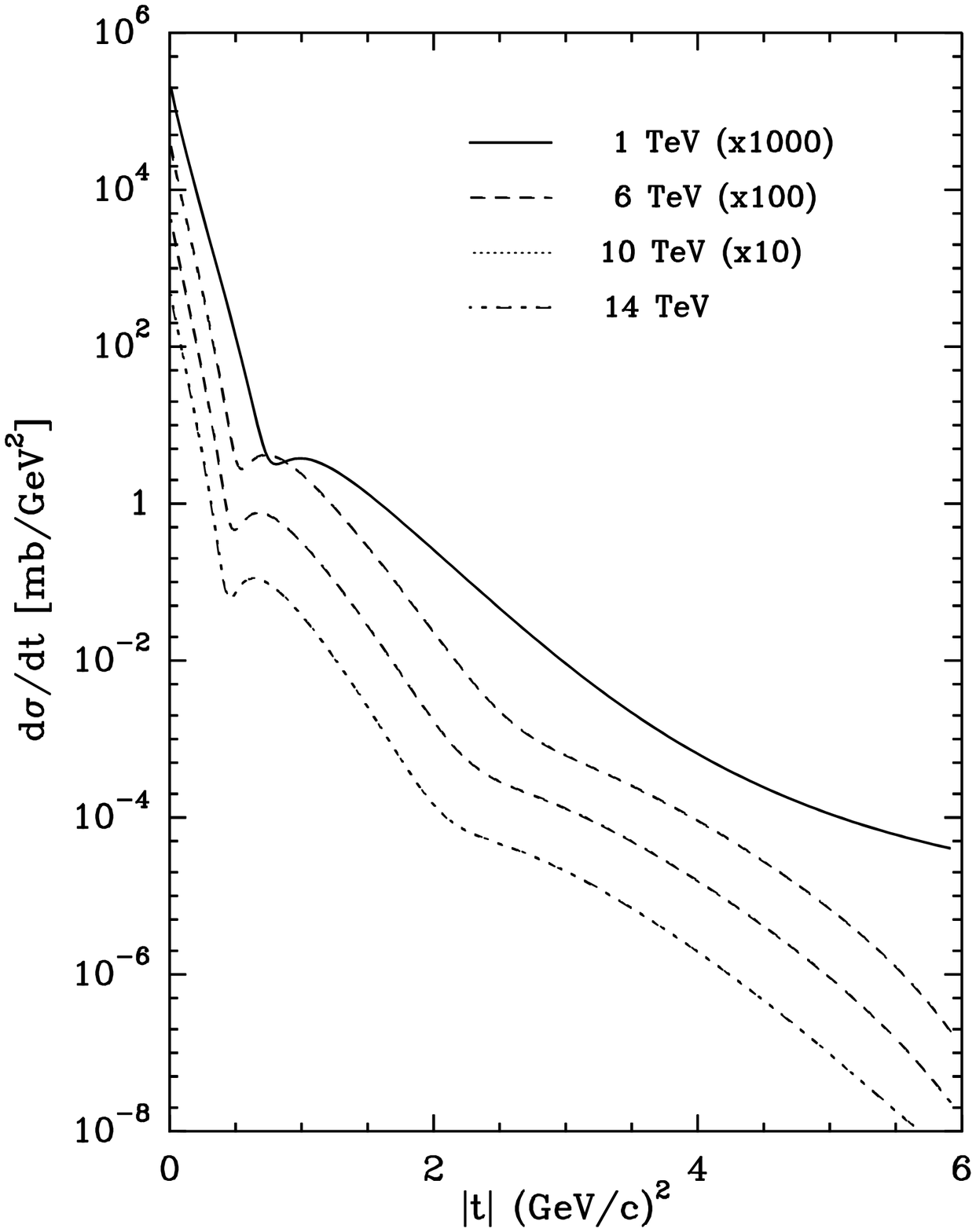,height=8.7cm,width=6.6cm}
\end{minipage}
\vspace*{-2.0ex}
\caption{
The ratio $\sigma_{el}/ \sigma_{tot}$  as function of $\sqrt{s}$ (left)
data from \cite{r17},
the differential cross section  for different energy bins as a function
of the momentum transfer (right).}
\label{fi:fig2}
\end{figure}

For completeness, in order to describe the very small $t$-region, one should 
add to the hadronic amplitude considered above, the Coulomb amplitude 
whose expression is 
$a^{C}(s,t) = 2 \alpha [s/|t|]G_{em}^2(t)  exp[\pm i\alpha \phi(t)]$, 
where $\alpha$ is the fine structure constant, $G_{em}(t)$ is the 
electromagnetic form factor, $\phi(t)$ is the West-Yennie phase \cite{WY} 
and the $\pm$ sign corresponds to $p p$ and $\bar p p$.
With the present parameters we obtained a $\chi^2/pt$=1840/431,
in comparison with the same data and the 1978 parameters $\chi^2$=4275.
In Fig. \ref{fi:fig1}, the variation of the total cross section and the
ratio $\rho = \mbox{Re}~a(s,0)/ \mbox{Im}~a(s,0)$ 
is shown as a function of the energy.
In Fig. \ref{fi:fig2}, the ratio $\sigma_{el}/\sigma_{tot}$ is an 
increasing function of the energy, the differential cross section as a 
function of the momentum transfer is plotted for different bins of energy
in the LHC energy domain.
In particular, at the LHC nominal energy,
$\sqrt{s} = 14$TeV, we predict $ \sigma_{tot} = 103.6\pm 1.1$mb,
$\rho = 0.122\pm 0.003$ and $\sigma_{el}/\sigma_{tot} = 0.274\pm 0.008$.
A significant departure from the previous $\rho$ value
would imply a violation of dispersion relations due to an unexpected
behavior of the scattering amplitude as discussed in \cite{r14}.

In conclusion,
after 30 years of existence the BSW impact picture 
remains reliable  to describe high energy hadron-hadron elastic scattering.
The structure of the pomeron amplitude we proposed has an
energy dependence deduced from the high energy behavior of tower
diagrams in QFT.
Over the years, with the advent of new experimental data, in particular,
at Tevtron and SPS,
we have obtained a more precise determination of the parameters
which leads to an improvement of the predictive power.
A crucial test of the high energy behavior of the impact picture
will be provided
by  ATLAS and TOTEM experiments at LHC\footnote{For a review see
\cite{r15}}.

\clearpage\newpage
\end{document}